\documentclass[prd,showpacs,preprintnumbers,superscriptaddress]{revtex4}
\usepackage{amssymb,hyperref}
\usepackage[dvips]{color}
\usepackage[dvips]{graphicx}

\newcommand{\bvec}[1]{\ensuremath{\mbox{\boldmath $\mathrm{#1}$}}}

\begin{document}

\title{Lattice Computations of the Pion Form Factor}

\author{Frederic D.\ R.\ Bonnet}
\email{fbonnet@jlab.org}
\affiliation{Department of Physics, University of Regina, Regina, SK,
  S4S 0A2, Canada}
\affiliation{Thomas Jefferson National Accelerator Facility, Newport News,
  VA 23606, USA}

\author{Robert G.\ Edwards}
\email{edwards@jlab.org}
\affiliation{Thomas Jefferson National Accelerator Facility, Newport News,
  VA 23606, USA}

\author{George T.\ Fleming}
\email{George.Fleming@Yale.edu}
\affiliation{Thomas Jefferson National Accelerator Facility, Newport News,
  VA 23606, USA}
\affiliation{Sloane Physics Laboratory, Yale University, New Haven,
  CT 06520, USA}

\author{Randy Lewis}
\email{Randy.Lewis@uregina.ca}
\affiliation{Department of Physics, University of Regina, Regina, SK,
  S4S 0A2, Canada}

\author{David G.\ Richards}
\email{dgr@jlab.org}
\affiliation{Thomas Jefferson National Accelerator Facility, Newport News,
  VA 23606, USA}

\collaboration{Lattice Hadron Physics Collaboration}
\noaffiliation

\begin{abstract}
We report on a program to compute the electromagnetic form factors of
mesons.  We discuss the techniques used to compute the pion form
factor and present results computed with domain wall valence fermions
on MILC \texttt{asqtad} lattices, as well as with Wilson fermions on
quenched lattices.  The methods can easily be extended to
$\rho\to\gamma\pi$ transition form factors.
\end{abstract}

\pacs{13.40.Gp,14.40.Aq,12.38.Gc}

\maketitle 

\section{\label{sec:introduction}INTRODUCTION}

The pion electromagnetic form factor is often considered a good
observable for studying the onset, with increasing energy, of the
perturbative QCD regime for exclusive processes.  It is believed that,
because the pion is the lightest and simplest hadron, a perturbative
description will be valid at lower energy scales than predictions for
heavier and more complicated hadrons such as the nucleon
\cite{Isgur:1984jm}.

\begin{figure}[ht]
\includegraphics[width=\textwidth]{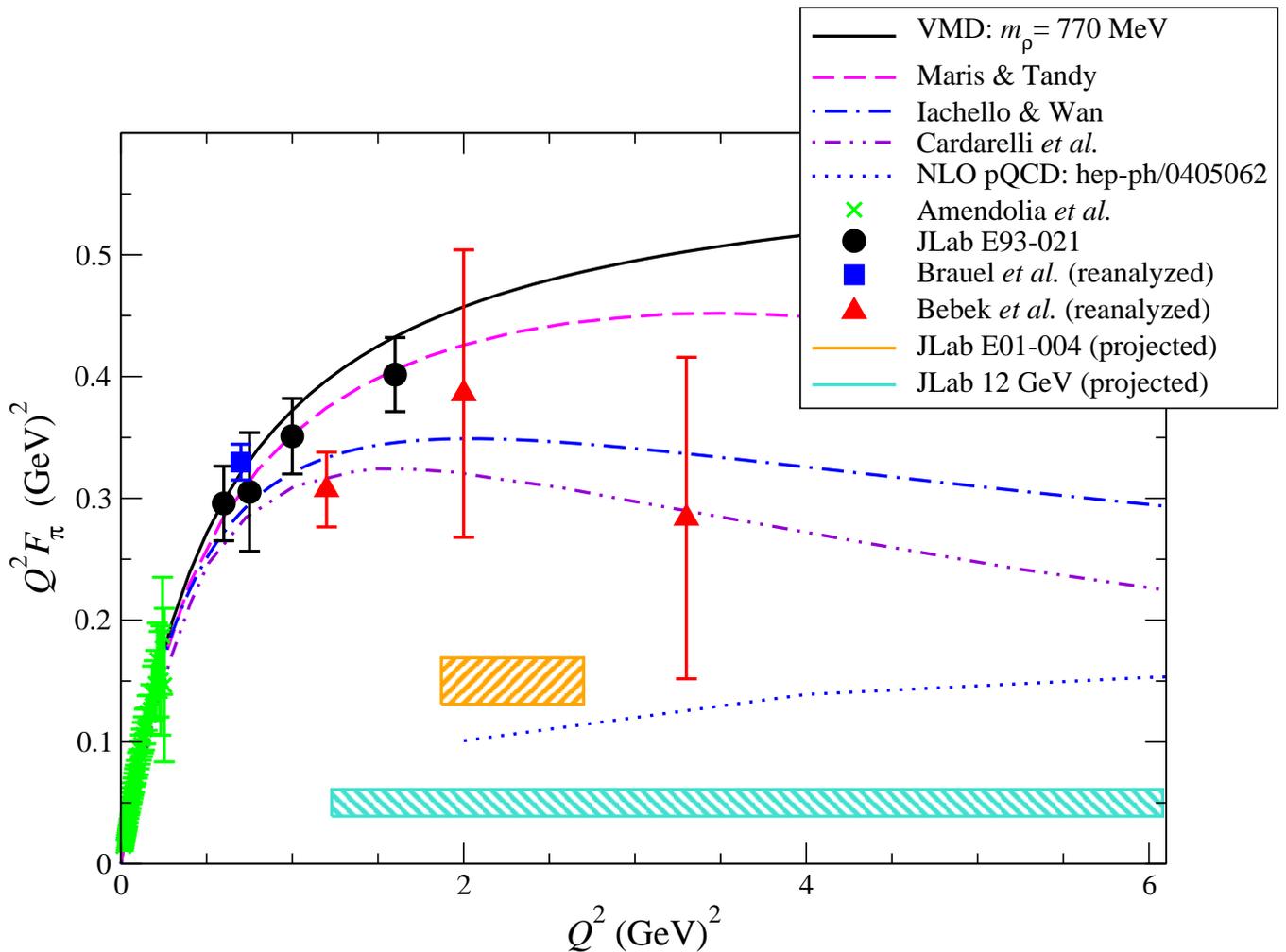}
\caption{\label{fig:Fpi_expt}Summary of experimental data
for the pion electromagnetic form factor; shaded regions indicate
expected sensitivities and coverage of future results.  The lines
indicate theoretical calculations of the form factor, as described in
the text.}
\end{figure}

A pseudoscalar particle has only a single electromagnetic form factor,
$F(Q^2)$, where $Q^2$ is the four-momentum transfer, and furthermore
at $Q^2 = 0$, this form factor is normalized to the electric charge of
the particle, $F(Q^2 = 0) = 1$; the magnetic form factor vanishes.
Thus in this paper, we will be measuring the form factor of the
positively charged $\pi^+$.  The experimentally observed behavior of
the form factor at small momentum transfer is well described by the
vector meson dominance (VMD) hypothesis
\cite{Holladay:1955,Frazer:1959gy,Frazer:1959}
\begin{equation}
\label{eq:vmd_form}
F_\pi(Q^2) \approx \frac{1}{1+Q^2 \left/ m_\mathrm{VMD}^2 \right.}
\quad \mathrm{for} \quad Q^2 \ll m_\mathrm{VMD}^2.
\end{equation}
The current experimental situation is presented in
Fig.~\ref{fig:Fpi_expt}.

For $Q^2\le 0.28\ \mathrm{GeV}^2$, Amendolia
\textit{et al.}~\cite{Amendolia:1986wj} have determined the form factor
to high precision from scattering of very high energy pions from
atomic electrons in a fixed target.  For higher $Q^2$, the form factor
has been determined from quasi-elastic scattering from a virtual pion
in the proton \cite{Bebek:1978pe,Brauel:1979zk,Volmer:2000ek}.  In this case,
the extracted values for the form factor must depend upon some theoretical
model \cite{Gutbrod:1972qr,Guidal:1997hy,Vanderhaeghen:1998ts}
for extrapolating the observed scattering from virtual pions
to the expected scattering from on-shell pions.  As the models have become
more sophisticated, the earlier data \cite{Bebek:1978pe,Brauel:1979zk}
have been reanalyzed \cite{Blok:2002ew} for consistency.
Shown in Fig.~\ref{fig:Fpi_expt} are the results of some model
calculations which seem to cover the range of existing predictions
\cite{Maris:2000sk,Cardarelli:1995dc,Iachello:2004ki,Iachello:2004}.

Given the dominance of the rho meson resonance in the time-like region
$(Q^2 < 0)$ of the pion form factor, perhaps it is not too surprising
that the space-like form factor is well described at low $Q^2 > 0$
by a VMD-inspired monopole form with only a contribution
from the lightest vector resonance
($m_\mathrm{VMD} \sim m_\rho \approx 0.77\ \mathrm{GeV}$).
What is striking is that it accurately describes all experimental data
even up to scales of $Q^2 \gtrsim 1\ \mathrm{GeV}^2$.
Furthermore, VMD predicts that the form factor should scale
as $F_\pi(Q^2) \sim 1 / Q^2$ for $Q^2 \gg m_\mathrm{VMD}^2$, the same scaling
predicted at asymptotically high $Q^2$ in perturbative QCD
\cite{Brodsky:1973kr,Brodsky:1975vy}.
One crucial fact which makes the pion form factor an ideal observable
for studying the interplay between perturbative and non-perturbative
QCD is that its asymptotic normalization can be determined
from pion decay\cite{Radyushkin:1977gp,Efremov:1978,Efremov:1978rn,
Efremov:1979qk,Jackson:1977,Farrar:1979aw,Lepage:1979zb}
\begin{equation}
F_\pi(Q^2) = \frac{8\pi\alpha_s(Q^2)f_\pi^2}{Q^2} \quad \mathrm{as} \quad
Q^2 \to \infty .
\end{equation}
Higher order perturbative calculations of the hard contribution to the form
factor \cite{Stefanis:1998dg,Stefanis:1998ud,Stefanis:2000vd,Bakulev:2004cu}
do not vary significantly from this value,
and are shown in Fig.~\ref{fig:Fpi_expt}.
At the largest energy scale where reliable experimental measurements have
so far been obtained, around $Q^2 \simeq 2\ \mathrm{GeV}^2$,
the data are 100\% larger than this pQCD asymptotic prediction.

This situation raises many questions.  At what scale does the form factor
vary with $Q^2$ as predicted by pQCD?  However, merely observing the
proper $Q^2$ dependence is insufficient as Eq.~(\ref{eq:vmd_form})
has the same asymptotic $Q^2$ dependence as pQCD but is numerically
about twice as large.  How rapidly will the data approach the pQCD prediction
and at what scale will pQCD finally agree with the data?  These are questions
which Lattice QCD calculations are ideally suited to address, provided
we can get reliable results for momentum transfer on the order
of a few to several $\mathrm{GeV}^2$.

Early lattice calculations validated the vector meson dominance hypothesis
at low $Q^2$ \cite{Martinelli:1988bh,Draper:1989bp}. Recent lattice results
\cite{vanderHeide:2003ip,vanderHeide:2003kh,vanderHeide:2004rj,Nemoto:2003ng,
Abdel-Rehim:2004sp, Abdel-Rehim:2004}, including some of our own preliminary
results \cite{Bonnet:2003pf,Bonnet:2003aa}, have somewhat extended the range
of momentum transfer, up to $2\ \mathrm{GeV}^2$, and the results remain
consistent with VMD and the experimental data.

\section{\label{sec:techniques}LATTICE COMPUTATION OF $F_\pi(Q^2)$}

\begin{figure}[ht]
  \includegraphics[width=0.5\textwidth]{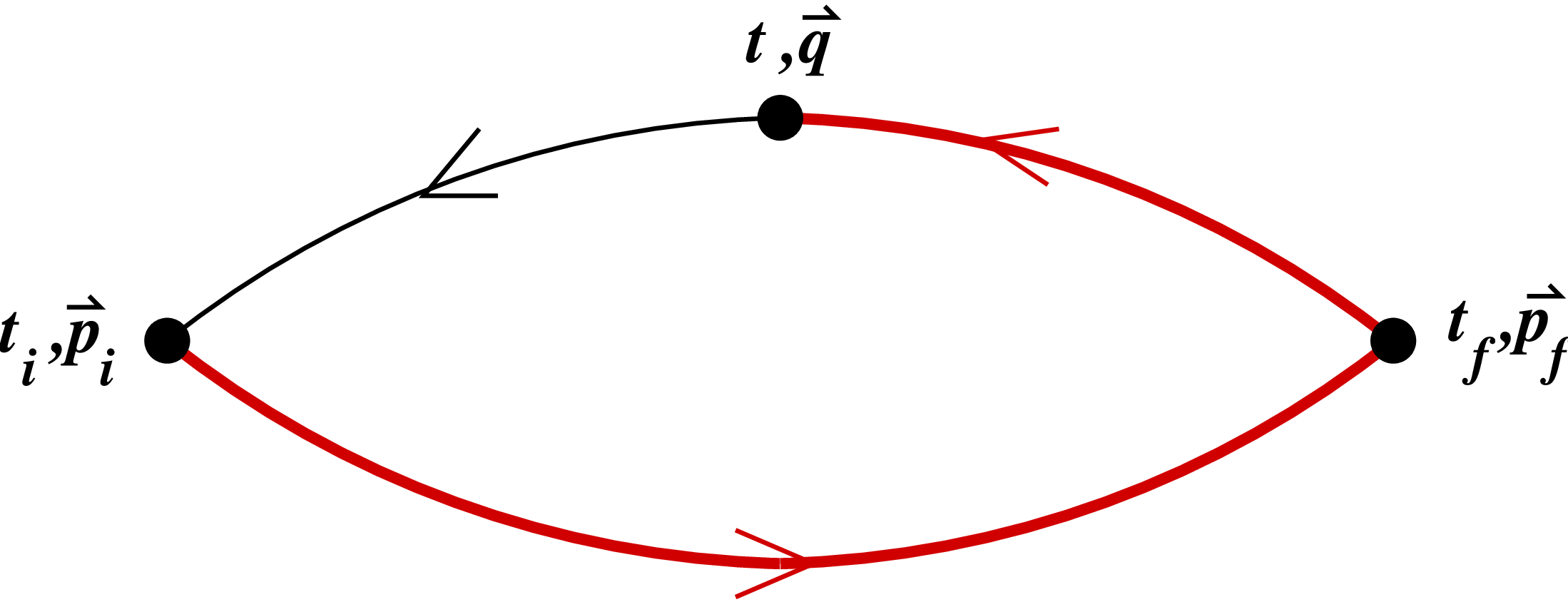}
  \caption{
    \label{fig:threepoint}
    The quark propagators used to compute the pion form factor.
  }
\end{figure}

The electromagnetic form factor is obtained in lattice QCD simulations
by placing a charged pion creation operator at Euclidean time $t_i$,
a charged pion annihilation operator at $t_f$ and a vector current insertion
at $t$ as shown in Fig.~\ref{fig:threepoint}.  A standard quark propagator
calculation provides the two propagator lines that originate from $t_i$.
The remaining quark propagator, originating from $t_f$ is obtained
via the \textit{sequential source method}:  (1) completely specify
the quantum numbers, including momentum $\bvec{p}_f$,
of the annihilation operator to be placed at $t_f$ and (2)
contract the propagator from $t_i$ to $t_f$ to the annihilation operator
and use that product as the source vector of a second, sequential
propagator inversion.  The resulting sequential propagator appears
as the thick line in Fig.~\ref{fig:threepoint} extending from $t_i$
to $t$ via $t_f$.  Given these two propagators, the diagram can be computed
for all possible values of insertion position $t$ and insertion
momenta $\bvec{q}$; the initial momentum $\bvec{p}_i$ is determined by
momentum conservation $\bvec{p}_i  = \bvec{p}_f - \bvec{q}$.

Furthermore, with the same set of propagators, any current can be inserted
at $t$ and any meson creation operator can be contracted at $t_i$.
So, the diagram relevant to determining the form factor
for the transition $\rho^+ \to \gamma^* \pi^+$ can be computed without
further quark propagator calculations.  By applying the sequential source
method at the sink, the trade-off is that the entire set of sequential
propagators must be recomputed each time new quantum numbers are needed
at the sink, particularly $\bvec{p}_f$.

We can extract the pion energies $E_\pi(\bvec{p})$ using standard lattice
techniques of fitting pion correlation functions from which we can compute
the momentum transfer
\begin{equation}
-Q^2 = \left[ E_\pi(\bvec{p}_f) - E_\pi(\bvec{p}_i) \right]^2
      - ( \bvec{p}_f - \bvec{p}_i )^2
\end{equation}
which should be non-positive if the pion spectral function is
well-behaved.  Since the largest $Q^2$ for a given $\left|\bvec{p}\right|^2$
occurs in the Breit frame, $\bvec{p}_f = -\bvec{p}_i$, it is important to
choose a non-zero $\bvec{p}_f$ to achieve large momentum transfer;
indeed $\bvec{p}_f = \bvec{0}$ yields a vanishing $Q^2$ for all
$\bvec{q}$ in the chiral limit.

The form factor, $F(Q^2)$, is defined by
\begin{equation}
\left\langle
  \pi(\bvec{p}_f) \left| V_\mu(0) \right| \pi(\bvec{p}_i)
\right\rangle_\mathrm{continuum} = Z_V \left\langle
  \pi(\bvec{p}_f)\left| V^\mathrm{lat}_\mu(0) \right| \pi(\bvec{p}_i)
\right\rangle = F(Q^2)(p_i+p_f)_\mu
\end{equation}
where $V^\mathrm{lat}_\mu(x)$ is a particular lattice discretization of
the continuum vector current.  In this work, local and point-split
vector currents were used:
\begin{eqnarray}
V_\mu^\mathrm{loc}(x) & = & \overline{\psi}(x) \gamma_\mu \psi(x) \\
V_\mu^\mathrm{p.s.}(x) & = & \frac{1}{2} \left\{
  \overline{\psi}(x+\widehat{\mu}) U_\mu^\dagger(x)
    \left[ 1 + \gamma_\mu \right] \psi(x)
  - \overline{\psi}(x) U_\mu(x) \left[ 1 - \gamma_\mu \right]
    \psi(x+\widehat{\mu})
\right\} .
\end{eqnarray}
$Z_V$ is the corresponding matching factor between the lattice current
and continuum; for Wilson fermions, the point-split current is conserved
on the lattice:  $Z_V^\mathrm{p.s.} =1$ (Wilson).  The three-point correlation
function depicted in Fig.~\ref{fig:threepoint} is given by
\begin{equation}
\Gamma_{\pi\mu\pi}^{AB}(t_i,t,t_f,\bvec{p}_i,\bvec{p}_f) =
\sum_{\bvec{x}_i,\bvec{x}_f}
e^{-i(\bvec{x}_f-\bvec{x})\cdot\bvec{p}_f}
\ e^{-i(\bvec{x}-\bvec{x}_i)\cdot\bvec{p}_i} \left\langle 0\left|
  \phi_B(x_f) V_\mu(x) \phi_A^\dagger(x_i)
\right| 0 \right\rangle
\end{equation}
where $\phi_A^\dagger(x)$ and $\phi_B(x)$ are creation and annihilation
operators with pion quantum numbers.  The $A$ and $B$ indicate that different
operators may be used at the source and sink,
\textit{i.e.}\ smeared source and point sink or pseudoscalar source
and axial vector sink.
 
Inserting complete sets of hadron states, and requiring $t_i \ll t \ll t_f$,
gives
\begin{equation}
\Gamma_{\pi\mu\pi}^{AB}(t_i,t,t_f,\bvec{p}_i,\bvec{p}_f)
\to \left\langle 0 \left| \phi_B(0) \right| \pi(\bvec{p}_f)\right\rangle
\frac{e^{-(t_f-t)E_\pi(\bvec{p}_f)}}{2 E_\pi(\bvec{p}_f)}
\left\langle \pi(\bvec{p}_f) \left|
  V_\mu(0)
\right| \pi(\bvec{p}_i) \right\rangle
\frac{e^{-(t-t_i)E_\pi(\bvec{p}_i)}}{2 E_\pi(\bvec{p}_i)}
\left\langle \pi(\bvec{p}_i) \left|
  \phi_A^\dagger(0)
\right| 0 \right\rangle.
\end{equation}
Similarly for the two-point correlator, with $t_i \ll t_f$,
\begin{equation}
\Gamma_{\pi\pi}^{AB}(t_i,t_f,\bvec{p}) \to
\left\langle 0 \left| \phi_B(0) \right| \pi(\bvec{p}) \right\rangle
\frac{e^{-(t_f-t_i)E_\pi(\bvec{p})}}{2 E_\pi(\bvec{p})}
\left\langle \pi(\bvec{p}) \left| \phi_A^\dagger(0) \right| 0 \right\rangle.
\end{equation}

We use both the pseudoscalar density, $\phi^{(1)}(x) =
\overline\psi(x) \gamma_5 \psi(x)$, and the temporal component of the
axial vector current, $\phi^{(2)}(x) = \overline\psi(x) \gamma_5
\gamma_4 \psi(x)$, as pion interpolating operators, constructed from
both local and smeared quark fields, denoted by L and S
respectively.  We adopt gauge-invariant Gaussian smearing
\begin{equation}
b(x) \to \left(
1 + \frac{\omega}{N} \bvec{\nabla} U
\right)^N b(x),
\end{equation}
where $\omega$ and $N$ are the tunable parameters used to specify the
smearing radius; the flavor structure is suppressed. The respective
merits of these interpolating operators will be discussed later.  For these
operators, we define the following amplitudes
\begin{eqnarray}
\label{eq:ZL1}
\left\langle 0 \left| \phi_L^{(1)}(0) \right| \pi(\bvec{p}) \right\rangle
& = & Z_L^{(1)} \\
\label{eq:ZS1}
\left\langle 0 \left| \phi_S^{(1)}(0) \right| \pi(\bvec{p}) \right\rangle
& = & Z_S^{(1)}(\left|\bvec{p}\right|) \\
\label{eq:ZL2}
\left\langle 0 \left| \phi_L^{(2)}(0) \right| \pi(\bvec{p}) \right\rangle
& = & Z_L^{(2)}(\left|\bvec{p}\right|) \\
\label{eq:ZS2}
\left\langle 0 \left| \phi_S^{(2)}(0) \right| \pi(\bvec{p}) \right\rangle
& = & Z_S^{(2)}(\left|\bvec{p}\right|).
\end{eqnarray}
The overlap of the operator $\phi_L^{(2)}$ has trivial $\bvec{p}$
dependence arising from the Lorentz structure of the operator
\begin{equation}
Z_L^{(2)} = E(\mid \bvec{p} \mid) f_{\pi}.
\end{equation}
However, the introduction of an additional three-dimensional scale
introduces non-trivial $\bvec{p}$ dependence for the smeared overlaps
$Z_S^{(1)}$ and $Z_S^{(2)}$.

We employ two methods to determine the form factor $F_\pi(Q^2)$.  The
first method, which we call the \textit{fitting method}, involves
a fit of the relevant two- and three-point functions to simultaneously
extract the form factor, the energies $E_\pi(\bvec{p})$ and the
amplitudes $Z(\bvec{p})$ in a single covariant, jackknifed fit.

The second method, which we call the \textit{ratio method},
starts by determining the energies $E_\pi(\bvec{p})$, either by fits to
the correlators at non-zero momentum or from a dispersion relation,
and then constructing the following ratio which is independent of
$Z_L^{(1)}$, $Z_S(|\bvec{p}|)$ and all Euclidean time exponentials for
sufficiently large temporal separations:
\begin{equation}
F(Q^2,t) = \frac{
  \Gamma_{\pi 4 \pi}^{AB}(t_i,t,t_f,\bvec{p}_i,\bvec{p}_f)
  \Gamma_{\pi\pi}^{CL}(t_i,t,\bvec{p}_f)
}{
  \Gamma_{\pi\pi}^{AL}(t_i,t,\bvec{p}_i)
  \Gamma_{\pi\pi}^{CB}(t_i,t_f,\bvec{p}_f)
} \left(
  \frac{2 Z_V E_\pi(\bvec{p}_f)}{E_\pi(\bvec{p}_i)+E_\pi(\bvec{p}_f)}
\right)
\end{equation}
where the indices $A$, $B$ and $C$ can be either $L$ (local) or $S$
(smeared).  As part of our program, we expect to determine the
relative merits of each extraction method.

\section{\label{sec:simulation}SIMULATION DETAILS}

Our first calculations were performed on quenched configurations
generated with the Wilson gauge action at $\beta=6.0$, corresponding
to $a^{-1} \approx 2\ \mathrm{GeV}$.  The propagators were computed
using the unimproved Wilson fermion action with Dirichlet boundary
conditions in the temporal direction, and periodic boundary conditions
in the spatial directions.  For Wilson fermions at these lattice
spacings, the exceptional configuration problem is rather mild
particularly when compared to that for the non-perturbatively improved clover
action.  This enabled us to reach pion masses of $300\ \mathrm{MeV}$ without
observing any exceptional configurations, whereas computations using
the clover action are limited to pion masses in excess of roughly
$400\ \mathrm{MeV}$ \cite{vanderHeide:2003ip} due to the systematic errors
caused by the frequent occurrence of exceptional configurations.
A detailed listing of the simulation parameters is provided
in Tab.~\ref{tab:Wilson_kappas}.  

\begin{table}[t]
  \caption{\label{tab:Wilson_kappas}Simulation details for quenched Wilson
    fermion calculations at $a^{-1} \approx 2\ \mathrm{GeV}$}.
  \begin{tabular}{ll|llll|c}
    $\kappa$ & volume         & $am_\rho$   & $am_\pi$    & $m_\pi/m_\rho$ &
      $m_\pi L$ & $Z_V^\mathrm{loc}$ \\
    \hline
    0.1480   & $16^3\times32$ & 0.7187(39)  & 0.6752(45)  & 0.943(10)      &
      10.80(7)  & 0.82725(51) \\
    0.1500   & $16^3\times32$ & 0.6387(57)  & 0.5854(45)  & 0.900(13)      &
       9.36(7)  & 0.78486(46) \\
    0.1520   & $16^3\times32$ & 0.5540(83)  & 0.4851(67)  & 0.876(14)      &
       7.76(11) & 0.74197(54) \\
    0.1540   & $16^3\times32$ & 0.4682(124) & 0.3752(73)  & 0.801(23)      &
       6.00(11) & 0.70459(39) \\
    \hline
    0.1555   & $24^3\times32$ & 0.4209(88)  & 0.2613(29)  & 0.621(14)      &
       6.27(7)  &             \\
    0.1563   & $24^3\times32$ & 0.4014(68)  & 0.1921(29)  & 0.479(10)      &
       4.61(7)  & 0.65676(43) \\
    \hline
    0.1566   & $32^3\times48$ & 0.3724(145) & 0.1629(36)  & 0.437(19)      &
       5.21(12) & 0.65553(14) \\
  \end{tabular}
\end{table}

The pion masses attainable in quenched domain-wall fermion (DWF)
calculations are limited only by finite volume effects and available
computing power.  So far, however, quenched DWF computations of the
pion form factor have only explored pion masses down to $390\ \mathrm{MeV}$,
at $a^{-1} \approx 1.3\ \mathrm{GeV}$~\cite{Nemoto:2003ng}.  An
important advantage of DWF fermions is that they are automatically
$\mathcal{O}(a)$-improved. An alternative approach using twisted-mass
QCD (tmQCD), in which Wilson fermion results are also $\mathcal{O}(a)$
improved with just double the effort\cite{Frezzotti:2003xj}, has been
explored separately\cite{Abdel-Rehim:2004sp}.

\begin{table}[ht]
  \caption{\label{eq:DWF_details}Simulation details for domain wall fermion
    calculations on 20$^3\times$64 dynamical MILC \texttt{asqtad} lattices
    at $a^{-1} \approx 1.6\ \mathrm{GeV}$.}
  \begin{tabular}{ll|l|ccll|cc}
    $am_{ud}$ & $am_s$ & $am_\mathrm{val}$ & $m_\rho\ \mathrm{(MeV)}$ &
      $m_\pi\ \mathrm{(MeV)}$ & $m_\pi/m_\rho$ & $m_\pi L$ &
      $Z_V^\mathrm{loc}$ & $Z_V^\mathrm{p.s.}$ \\
    \hline
    0.01 & 0.05 & 0.01  &  956(22) & 318(3) & 0.333(8)  & 3.97(4) &
      1.0714(55) & 1.1098(23) \\
    0.05 & 0.05 & 0.05  &  955(19) & 602(5) & 0.630(12) & 7.68(6) &
      1.0890(55) & 1.0835(37) \\
    0.05 & 0.05 & 0.081 & 1060(14) & 758(5) & 0.715(10) & 9.66(6) &
      1.1199(14) & 1.0833(13) \\
  \end{tabular}
\end{table}

The Lattice Hadron Physics Collaboration (LHPC) has been performing
unquenched hadron structure calculations using MILC $N_f=2+1$ and $N_f=3$
configurations generated with staggered \texttt{asqtad} sea quarks
\cite{Negele:2003ma,Schroers:2003mf,Negele:2004iu,Renner:2004ck,
Hagler:2004er}.  Valence propagators were computed using domain wall
quarks with domain wall height $m=1.7$ and extent $L_s=16$
of the extra dimension and Dirichlet boundary conditions imposed
32 timeslices apart. The MILC configurations were HYP blocked
\cite{Hasenfratz:2001hp} before the valence propagators were computed
to avoid unacceptably large residual chiral symmetry breaking.
For our pion form factor calculation we chose configurations
separated by at least 12 HMD trajectories.  In some
of the other LHPC calculations, if measurements were computed
on configurations separated by 6 HMD trajectories, opposite halves
of the lattice were used to minimize autocorrelations.
The simulation parameters for the DWF computation are provided
in Tab.~\ref{eq:DWF_details}. A detailed study of the physical properties
of light hadrons composed of staggered quarks computed on these lattices
has recently been completed \cite{Aubin:2004wf}.

\section{\label{sec:results}RESULTS}

\begin{figure*}[hbt]
  \begin{center}
    \includegraphics*[width=\textwidth]{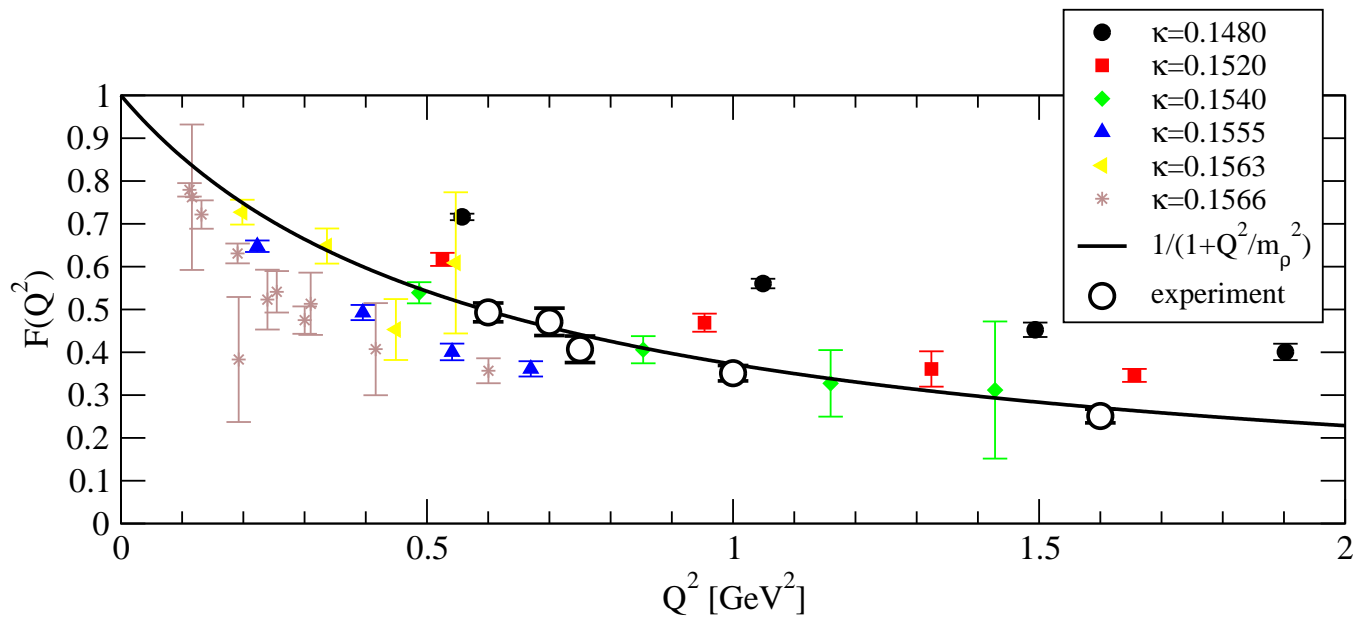}
  \end{center}
  \caption{\label{fig:Wilson_monopole}Results for the pion form factor
    as a function of $Q^2$ for each of the available $\kappa$ values,
    together with experimental determinations\protect\cite{Volmer:2000ek}.
    The curve shows the vector-meson-dominance prediction, using the
    experimental determination of the $\rho$ mass.}
\end{figure*}

Results for the quenched Wilson form factor computed by the fitting
method at each value of the quark mass are shown in
Fig.~\ref{fig:Wilson_monopole}.  Also indicated are experimental data
points and the vector meson dominance (VMD) prediction using the experimental
value for the $\rho$ meson mass.  While the data tend in the correct
direction with decreasing pion mass, the reader may notice that the form
factor for $300\ \mathrm{MeV}$ pions already lies below the physical curve.
This suggests that the pole mass that would be obtained from a
VMD fit to the lattice data would be considerably
lower than the physical $\rho$ meson mass.
This is perhaps expected since it is known that
$\mathcal{O}(a)$ scaling violations in the vector meson mass computed
with Wilson fermions yield an underestimate of the mass of roughly 20\%
\cite{Edwards:1998nh}, the same amount needed to move the form factor
points so as to lie above the continuum curve.

Because we would like to compute the pion form factor at large
momentum transfer, we have spent a substantial amount of effort on our
domain wall data set in extracting the pion energies and amplitudes at
relatively large momenta; the largest attainable momenta is
constrained by the fineness of the lattice spacing.  In the continuum
limit, the pion dispersion relation should approach the continuum one
\begin{equation}
E_\pi(\bvec{p})^2 = \bvec{p}^2 + E_\pi(0)^2.\label{eq:cont_disp}
\end{equation}
At a non-zero lattice spacing, a study of free lattice bosons suggests
the lattice dispersion relation
\begin{equation}
\label{eq:free_lattice_boson_dispersion}
\widehat{E}(\bvec{\widehat{p}})^2 = \bvec{\widehat{p}}^2 + \widehat{E}_\pi(0)^2
\end{equation}
where $\widehat{E}$ and $\widehat{p}$ are the ``lattice'' energy and
momentum respectively:
\begin{equation}
\widehat{E} = 2 \sinh \left( \frac{E}{2} \right) \qquad
\widehat{p}_x = 2 \sin  \left( \frac{p_x}{2} \right).
\end{equation}
Eqs.~(\ref{eq:cont_disp}) and
(\ref{eq:free_lattice_boson_dispersion}) agree in the small-momentum limit.

\begin{figure*}
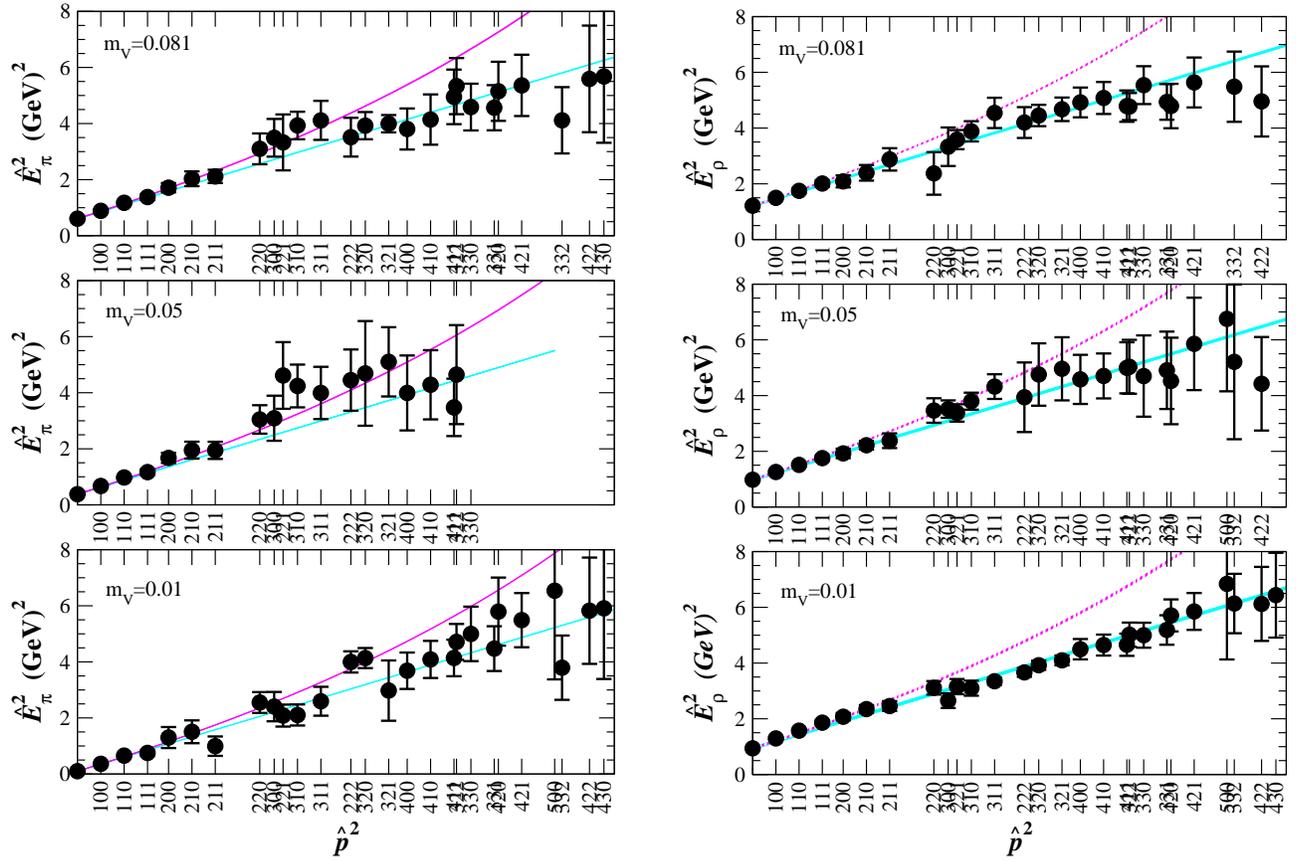

  \begin{center}
   
 \includegraphics[width=0.45\textwidth]{pion_disp.eps}
    \hspace{2em}
    \includegraphics[width=0.45\textwidth]{rho_disp.eps}
  \end{center}
  \caption{\label{fig:dwf_dispersion}Pion (left) and rho meson (right)
    dispersion relation \textit{vs.}\ continuum (upper) and lattice (lower)
    expectations curves.}
\end{figure*}
In Fig.~\ref{fig:dwf_dispersion}, we show the measured ``lattice''
energies against the lattice momenta at each of our quark masses,
together with curves representing the continuum and lattice dispersion
relations.  We see that both dispersion relations provide a reasonable
representation of the data, although there may be a slight flattening
of the data against the continuum curve at higher momenta.  These
results suggest that directly fitting all the data to either
dispersion relation, thereby reducing the number of fit parameters
needed to extract the form factor, may improve the relative signal to
noise of the remaining parameters.  This may help dramatically in the
ratio method, where the only fit parameters are the form factor and
the energies.

\begin{figure*}
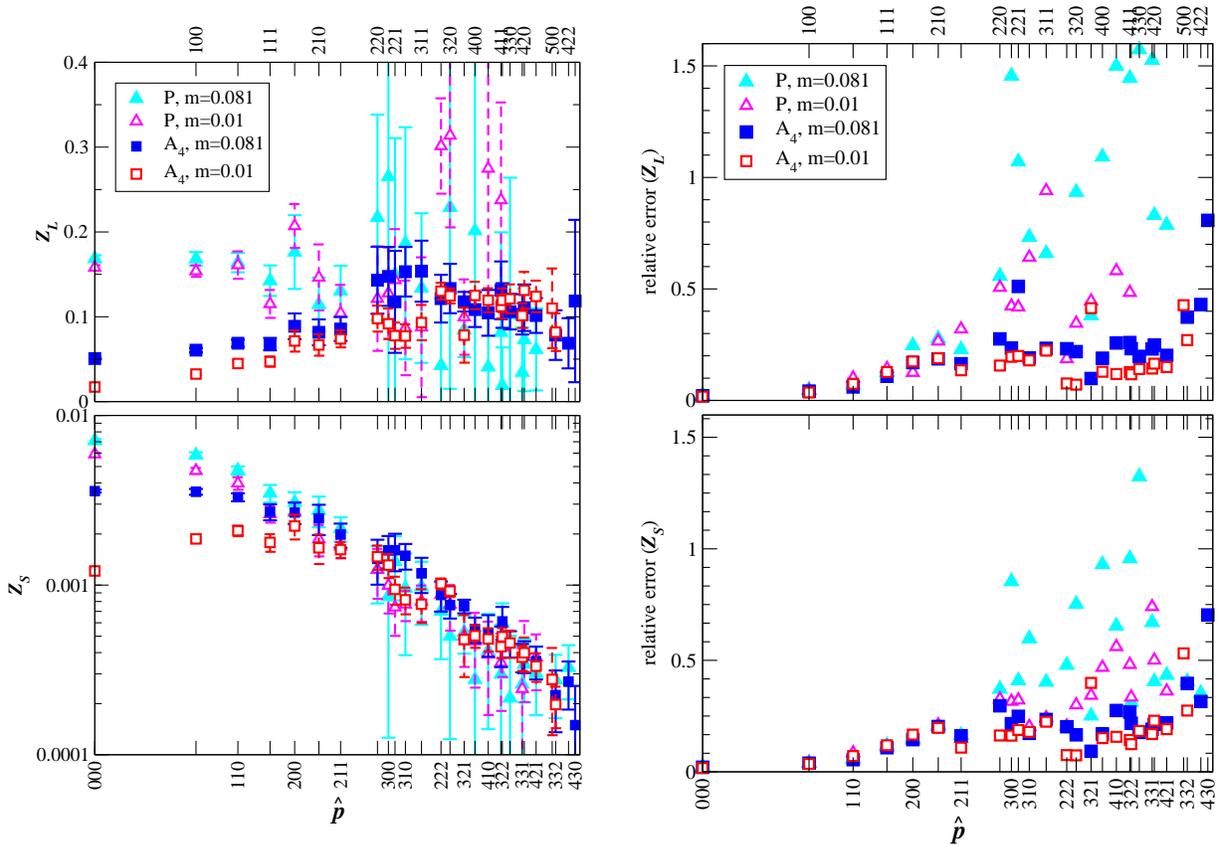

  \begin{center}
    \begin{minipage}{0.45\textwidth}
      \includegraphics[width=0.95\textwidth]{z_pion.eps}
    \end{minipage}
    \hspace{1ex}
    \begin{minipage}{0.45\textwidth}
      \includegraphics[width=0.95\textwidth]{z_pion_rel_err.eps}
    \end{minipage}
  \end{center}
  \caption{\label{fig:dwf_amplitudes}Results for the local (L) and smeared (S)
    pseudoscalar ({\large\color[named]{Cyan} $\blacktriangle$},
    {\large\color[named]{Magenta} \boldmath $\vartriangle$}) and axial vector
    ({\color[named]{Blue} $\blacksquare$},
    {\color[named]{Red} \boldmath $\square$}) pion source amplitudes
    $Z_A^{(i)}(\bvec{p})$ used in this study.}
\end{figure*}

In the fitting method, one must reliably extract not only the energies,
but the amplitudes, at high momenta. In Fig.~\ref{fig:dwf_amplitudes}
we present the four amplitudes of Eqs.~(\ref{eq:ZL1}) through (\ref{eq:ZS2})
that we estimate from the four two-point correlators we measure:
smeared-smeared and smeared-local for both pseudoscalar-pseudoscalar
and axial-axial operators.  In the fitting procedure, all four correlators
are constrained to have the same energy.  From the figure, we can see that
our expectations of $Z_L^{(1)} \propto \mathit{const}$
and $Z_L^{(2)}(\bvec{p}) \propto E_\pi(\bvec{p})$ are consistent
with the data.  We can also see from $Z_S^{(1)}(\bvec{p}=0)$ that the smeared
pseudoscalar operator has a strong overlap with the zero momentum pion
relative to the axial-vector operator.

We plot in Fig.~\ref{fig:dwf_amplitudes} just the relative error
estimates for the amplitudes.  This clearly demonstrates that
the statistical noise inherent in a particular source or sink operator
need not be correlated with the magnitude of the amplitude.
In particular, we see that beyond the few smallest momenta, the pseudoscalar
and axial-vector amplitudes are of the same order, but the inherent
noise of the pseudoscalar operator is unacceptably large for higher momenta.

\begin{figure}
\includegraphics[width=\textwidth]{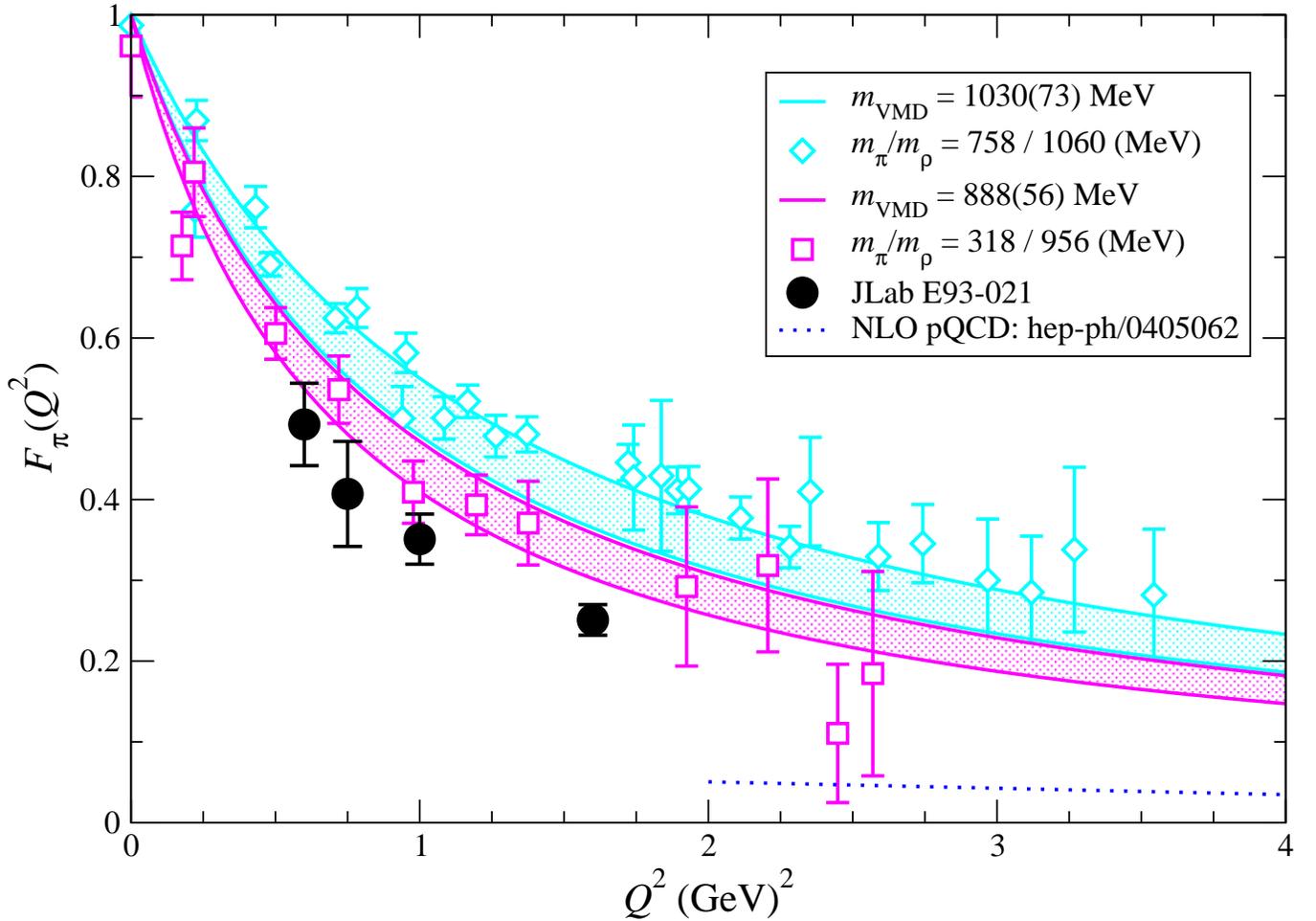}
\caption{\label{fig:F_px1_py0_pz0} Pion electromagnetic form factor
  for fixed sink momentum $\bvec{p}_f = (1,0,0)$ computed by the ratio method
  \cite{Bonnet:2003pf,Bonnet:2003aa} and imposing the lattice dispersion
  relation, Eq.~(\ref{eq:free_lattice_boson_dispersion}). Shaded regions
  are jackknife error bands for VMD fit.
}
\end{figure}

\begin{table}[ht]
  \caption{\label{tab:VMD_fits}Results of fit of form factor data
    to the VMD monopole \textit{ansatz} of Eq.~(\ref{eq:vmd_form}).}
  \begin{tabular}{l|c|c|c|c|c}
    $am_\mathrm{val}$ & $\bvec{p}_f$ & $m_\mathrm{VMD}$ (GeV) &
      $\chi^2/\mathrm{d.o.f.}$ & $\mathrm{d.o.f.}$ & $N_\mathrm{confs}$ \\
    \hline
    0.01  & (0,0,0) &    ---    &   ---    & --- &  51 \\
          & (1,0,0) & 0.888(56) & 1.37(72) & 13  & 251 \\
          & (1,1,0) & 0.30(20)  & 1.7(1.2) & 11  & 106 \\
    \hline
    0.05  & (0,0,0) & 1.278(87) & 2.8(1.3) & 20  & 104 \\
    \hline
    0.081 & (0,0,0) & 1.192(93) & 3(17)    & 15  &  49 \\
          & (1,0,0) & 1.030(73) & 2.1(1.7) & 22  &  70 \\
          & (1,1,0) & 1.022(87) & 3.6(2.2) & 23  &  73 \\
  \end{tabular}
\end{table}

Using the lattice dispersion relation and the ratio method described
above and in our previous work, we have computed in
Fig.~\ref{fig:F_px1_py0_pz0} the pion form factor for two of the
dynamical pion masses.  Fits of the data to the monopole form of
Eq.~(\ref{eq:vmd_form}) are also shown, where shaded regions
correspond to jackknife error bands and the central values for the
pole masses are given in the legend.  The data in
Fig.~\ref{fig:F_px1_py0_pz0} were computed with pseudoscalar pion sink
operator fixed at momentum $\bvec{p}_f=(1,0,0)$.
Tab.~\ref{tab:VMD_fits} summarizes the fits for all available
combinations of $m_\mathrm{val}$ and $\bvec{p}_f$.  Unconstrained linear
extrapolation of the two most reliable data points, $m_\mathrm{val}=0.01$
and 0.081 with $\bvec{p}_f=(1,1,0)$, gives an estimate
of $m_\mathrm{VMD} = 0.868(65)\ \mathrm{GeV}$ in the zero quark mass limit. 
From this we estimate the mean square charge radius of the pion to be
$\left\langle r^2 \right\rangle_\mathrm{VMD} = 0.310(46)\ \mathrm{fm}^2$,
which is significantly below the experimental value of
$\left\langle r^2 \right\rangle_\mathrm{expt} = 0.439(8)\ \mathrm{fm}^2$.
More dynamical quark masses, particularly lighter ones, will be needed
to perform a proper chiral extrapolation.

We were unable to compute statistically significant form factors with
the same pseudoscalar sink operator for $m_\mathrm{val}=0.01$ and
$\bvec{p}_f=(1,1,0)$ by the ratio method, apparently due to the poor
signal-to-noise inherent in the overlap of the chosen sink operator
with the higher momentum state.  For $m_\mathrm{val}=0.01$ and
$\bvec{p}_f=(0,0,0)$ the signal was rather better but 51 sequential
propagators were too few to allow for a stable fit to the VMD \textit{ansatz}.
We expect that an axial vector sink operator would be a better choice
for future calculations at non-zero sink momentum.  Some of us are also
extending our group-theoretical construction of extended baryon operators
\cite{Edwards:2003mv,Basak:2003nh,Basak:2003yd,Basak:2004hp,Basak:2004ib}
to include mesonic quantum numbers.

\section{\label{sec:Conclusions}CONCLUSIONS}

From our quenched Wilson form factor results, we find that both the
ratio method and the fitting method are useful tools for computing the
pion form factor.  Each method has different systematic errors, so the
extent to which both agree should give confidence that the systematic
errors are small and well understood.  A comparison of the Wilson form
factor results with the experimental pion form factor data suggests
that the former have large, and expected, discretization errors.

Domain-wall fermions are free of ${\cal O}(a)$ discretization
uncertainties to the extent that any residual mass is eliminated.
Thus our dynamical fermion results are obtained using domain-wall
fermions for the valence quarks computed on \texttt{asqtad}
lattices.  From a careful analysis of the pion spectrum, we recognize
the importance of using a large basis of pion operators so that we can
identify at least one operator at each pion momentum whose overlap
with the pion has a reasonable signal-to-noise ratio.  The local axial
vector operator appears to be a better choice than the pseudoscalar
operator for fitting higher momentum states since its overlap with a
given state increases in proportion with the pion energy and without
any degradation of the signal-to-noise ratio.  Our analysis enables us
to obtain the pion form factor for a pion mass approaching $300~{\rm
MeV}$ at a range of $Q^2$ commensurate with the current experimental program.

Given our existing analysis framework and the propagators and
sequential propagators that we have already computed, we can also
compute the $\rho^+\to\gamma\pi^+$ transition form factor.  We hope to
complete this analysis in the near future.

\section*{Acknowledgments}

This work was supported in part by the Natural Sciences and Engineering
Research Council of Canada and in part by DOE contract DE-AC05-84ER40150
Modification No. M175, under which the Southeastern Universities Research
Association (SURA) operates the Thomas Jefferson National Accelerator Facility.
Computations were performed on the 128-node and 256-node Pentium IV clusters
at JLab and on other resources at ORNL, under the auspices
of the National Computational Infrastructure for Lattice Gauge Theory,
a part of U.S.~DOE's SciDAC program.

\bibliography{main}

\end{document}